\documentclass[12pt,fleqn]{article}

\begin{document}
\title{Superoscillations and tunneling times}
\author{Yakir Aharonov$^{a,b}$, Noam Erez$^{a}$, and Benni Reznik$^{a}$
 {\ }\\
\small a) {\em \small School of Physics and Astronomy, Tel Aviv
University, Tel Aviv 69978, Israel}\\ \small b) {\em \small Department
of Physics, University of South Carolina, Columbia, SC 29208}
}
\date{October 16, 2001}
\maketitle

\begin{abstract}
It is proposed that superoscillations play an important role in
the interferences which give rise to superluminal effects.  To
exemplify that, we consider a toy model which allows for a wave
packet to travel, in zero time and negligible distortion a
distance arbitrarily larger than the width of the wave packet. The
peak is shown to result from a superoscillatory superposition at
the tail. Similar reasoning applies to the dwell time.
\end{abstract}

\section{Introduction}

Superluminal effects have been predicted in conjunction with
various quantum systems propagating in a forbidden zone.
 In these regimes, the probability density current is imaginary,
 and various operational definitions of the velocity of a wave packet
have been proposed -
in many examples, giving differing values.
In recent years, a number of experiments with superluminal photons
have been performed, reviving interest in the problem, as well as
controversy. The theoretical predictions have been verified - in
fact, different tunneling times have been measured in accordance
with the different operational definitions appropriate for the
various experimental set-ups.
(An excellent
review on the experimental situation, as well as the theoretical
background is afforded by \cite{CS:rev}.)

Causality is not directly violated. The signal
velocity (the velocity of propagation of an \emph{abrupt}
disturbance) is always subluminal. Other "velocities" may well be
superluminal, for example the group velocity of a wave packet. In
the latter case, the (local) peak of the packet appears at a point
where constructive interference builds up, much earlier than the
arrival of a freely propagating wave function. Thus the information
stored in the tail may be travelling way ahead of the peak and can
possibly be used to anticipate it.

The purpose of this article is to investigate further the nature
of the interferences that give rise to superluminality. It has
been noted by Steinberg \cite{Steinberg1,Steinberg2} that the
superluminality phenomena is associated with post-selection: for
instance, from a sample of particles that scatters of a barrier we
examine only the subsample that tunnel through. When a pre- and
post-selected system is subjected to a non-disturbing, "weak"
measurement, the outcome of the measurement, known as the "weak
value", can attain values that lie outside the spectrum of
eigenvalues of the measured observable\cite{YA:weak4}. Weak values
may hence be naturally related to the superluminal phenomena, as
indeed, Steinberg has already argued that the dwell time is a weak
value of a projector to the tunneling domain. The appearance of
unusual weak values has been associated with a unique interference
structure \cite{aharonov}, for which Berry \cite{berry} coined the
term "{\em superoscillations}". As an instructive example of a
superoscillatory function $F(k)$ consider
\begin{equation}
F(k; N, L) = \biggl[
         \biggl({1-L/x_0\over2}\biggr)e^{i k x_0 /N}
         +\biggl( {1+L/x_0\over2}\biggr)
         e^{-i k x_0 /N}
                \biggr]^N.
\label{func}
\end{equation}
Here, $N>1$ is an integer, and $L$ and $x_0$ being the
super and reference shifts.
For small $k$ we expand $\exp(ik x_0 /N)$ and find:
\begin{equation}
F(k; N, L) =e^{-ik L} \biggl[   1+   {(L^{2}-x_0^2)k^2\over2N} +
O (N^{-2})  \biggr]   \cong  e^{-ikL }.
\end{equation}
Although this function is a superposition of waves $e^{ikx}$ with
$|x|\le x_0$, in the interval $|k|<<
\sqrt{N}/\sqrt{L^{2}-x_0^2}\equiv \Delta k$, $F(k)$ behaves nearly
as a pure wave $e^{ikL}$ with $L$ arbitrary larger than $x_0$. In
the regime $|k|<\Delta k$ the function oscillates rapidly. The
number of these ``superoscillations'' is $\sim \sqrt N$. This
remarkable feature is derived at the expense of having the
function grow exponentially in other regions. In the example
above, for $|k|>\Delta k$, we get $F\sim e^N$.

In this article we will suggest that at least for certain cases,
the constructive interference giving rise to superluminal effects,
originates from a similar structure of superoscillations. To
exemplify that, we consider a toy model, (which extends on a
previous proposal of Olkhovsky Recami and Salesi
\cite{ORS:2barr}), which allows for a wave packet to traverse, in
a vanishing time and negligible distortion (the transmitted wave
packet is the first derivative of the incoming packet), a distance {\em
arbitrarily larger than the original size of the wave packet}. Hence
the peak is here reconstructed from the exponentially small tail
of the wave function. As far as we know, in the examples
discussed to date, the superluminal shift of the wave packet is
restricted. It is comparable to or much smaller than the initial
wave packet size. We then show that the resulting null group-delay
and dwell times\footnote{We follow the terminology of
\cite{CS:rev}. The group delay is defined as \mbox{$\tau_g \equiv
\hbar \frac{\partial}{\partial E} \arg (t) $}, where $t$ is the
transmission coefficient for tunneling through the region.
Sometimes however, the same name is used for the difference
between this value and the time it would take the same packet to
traverse an equal distance in free propagation (i.e. for the
additional delay introduced by the barrier). Other names for it
are: phase- Wigner- and stationary phase- time. In the
following we refer to the forward conditional dwell
time\cite{Steinberg1,Steinberg2}, as the dwell time.} Finally,
coming back to the role of post-selection we provide a rigorous
proof for Steinberg's claim \cite{Steinberg1,Steinberg2} that the delay time is a weak
value of a projector operator.

The article proceeds as follows. In section 2 we calculate the
dwell time and the group delay time for tunneling through $n$
delta-function barriers. In the low energy limit, both turn out to
be zero. We also derive the condition for the calculations to
apply for a wave-packet. Using this condition, we see that for
this system, the (negative) delay can be larger than the
uncertainty associated with the length of the wave-packet. Section
3 deals with the relation between superluminality and interference
effects in the tail of the wave function, and superoscillations.
The applicability to the example of section 2, of the explanations
of superluminality given in other cases, is discussed. Finally, in section
4 we elaborate on Steinberg's claim that the dwell time is a weak
value.



\section{The group delay and dwell
time for a particle tunneling through an $n$-delta-function potential}

Olkhovsky Recami and Salesi \cite{ORS:2barr} showed that a
Schr\"{o}dinger particle tunneling through a double rectangular
barrier traversed the distance between the bumps instantaneously
in the limit that its kinetic energy was much smaller than the
height of the barrier. Unlike previous examples of superluminal
tunneling, the length of the region of superluminality consists of
an arbitrarily long portion with zero potential, between the
bumps. Replacing the rectangular barriers in the example discussed
in \cite{ORS:2barr} by delta-function potentials, the calculations
can be made somewhat simpler, and are easily generalized to $n$
arbitrary delta bumps (still using the approximation of low
kinetic energy).

In this section we make a direct calculation of the transmission
coefficient for the stationary scattering of a scalar particle
obeying the Schr\"{o}dinger equation, off a multiple
delta-function potential. The time independent equation is the
same as for the scalar relativistic wave equation, and we focus on
the Schr\"{o}dinger equation merely for a simple concrete
interpretation.

\subsection {Transmission through a multiple delta-function potential.}
Consider the Schr\"{o}dinger equation with the following
potential:
\begin{equation} V(x) = \Sigma \alpha_{i} \delta (x-L_{i});\  L_{0} = 0.
\label{potential}
\end{equation}
The energy eigenfunctions have the form (for $x<0$ and $x>L_n$):
\begin{equation} \psi (x) = \left\{ \begin{array}{cc} Ae^{ikx}+Be^{-ikx} & x<0 \\
                                       Ce^{ikx}+De^{-ikx} & x>L_{n}  \end{array}
\right. \label{asympt}
\end{equation}
The coefficients satisfy:
\begin{equation}
 \left( \begin{array}{c}
            A \\ B \end{array}  \right) =
 M
 \left( \begin{array}{c} C \\ D  \end{array} \right),
\end{equation}

\begin{equation} M =  \prod_{i=1}^{n} \left[ \beta_{i} \ \left(
\begin{array}{cc}
1 & e^{-2ikL_{i}} \\ -e^{2ikL_{i}} & -1
\end{array}
\right) + I \right],\ \beta_j = \frac{m\alpha_j}{ik}
\label{tmatrix}
\end{equation}
In the limit of small kinetic energy ($|\beta_{i}| \gg 1$), we can
drop the $I$ matrices, as long as $n<\beta_i$. It is then
straightforward to prove by induction on $n$ that:

\begin{equation} M = \prod_{1}^{n} \beta_{i} \left( \begin{array}{cc}
\prod_{i=2}^{n} (1-z_{i}) & \prod_{i=2}^{n} (z_{i}^{-1}-1) \\
-\prod_{i=2}^{n} (1-z_{i}) & -\prod_{i=2}^{n} (z_{i}^{-1}-1)
\end{array}
\right) + O(1)
\end{equation}
where $z_{1}=1, z_{i} = e^{2ik(L_{i}-L_{i-1})} \ (i=2..n)$

As usual, we examine the case of ``stationary scattering''. To get
the (amplitude) transmission coefficient for probability current
flowing from the left,$t$, we put $A=1, B=r, C=t, D=0$ in eq.
(\ref{asympt}):

\begin{equation} \psi (x) = \left\{ \begin{array}{cc} e^{ikx}+re^{-ikx} & x<0
\\ t e^{ikx} & x>L_{n}  \end{array} \right. \label{eqtrans}
\end{equation}
and we see that $t =  M_{11}^{-1}$, so:

\begin{eqnarray} t = M_{11}^{-1} \approx \frac {\beta_{1}^{-1} \cdots \beta_{n}^{-1}}
{\prod_{i=2}^{n}(1-z_{i})} = \beta_{1}^{-1} \cdots \beta_{n}^{-1}
\frac {\prod_{i=2}^{n} z_{i}^{-\frac{1}{2}}}
{\prod_{i=2}^{n}(z_{i}^{-\frac{1}{2}}-z_{i}^{\frac{1}{2}})}
\nonumber \\  = \frac{\prod\beta_i^{-1}}{(-2i)^{n-1}}
\frac{e^{-ikL_{n}}}{\prod_{i=2}^{n}\sin(k(L_{i}-L_{i-1}))}
\label{tforn}
\end{eqnarray}
The stationary phase formula for the delay time, $\tau_g$:

\begin{equation} \tau_g \equiv \hbar \frac{\partial}{\partial E} \arg (t),
\label{taug} \label{} \end{equation} yields the value $\tau_g =
-\frac{mL}{\hbar k}=-\frac{L}{v(k)} $ for the delay, which cancels
the time for a free particle, and we get an overall zero time for
tunneling. Since this is true for all $k$, it should be true for
an arbitrary wave packet, \emph{as long as the stationary phase
approximation holds.} The condition for that is derived in the
next subsection.

\subsection{The condition for superluminal tunneling of a packet}
Restated for wave packets, our results so far can be summarized
as:

\begin{equation} \Psi (x,t) = \left\{ \begin{array}{cc} \int g(k)\left(e^{ikx}+r(k)e^{-ikx}\right)e^{-i\omega(k) t}dk & x<0 \\
                                       \int (-ik)C(k)g(k)e^{ik(x-L_{n})}e^{-i\omega(k) t}dk & x>L_{n}  \end{array} \right.
,\end{equation}
and
\begin{equation}
C(k) =
\frac{\prod\beta_i^{-1}}{-ik(-2i)^{n-1}
\prod_{i=2}^{n}\sin(k(L_{i}-L_{i-1}))}
\end{equation}

When $\Delta k$ is sufficiently small, the diffusion can be
ignored and $C(k)$ can be considered constant (as will be shown
shortly). Then we can again separate out the time dependence of
the wave function and the spatial part can be written:

\begin{equation} \psi (x) = \left\{ \begin{array}{cc} \phi(x) & x<0 \\
                                       C\phi'(x-L_n) & x>L_{n}  \end{array} \right.
\end{equation}
where $\phi(x)$ in the two regions is related through analytic
continuation. \\If \mbox{$\phi (x) = R(x)e^{iS(x)}$} where $R(x)$
is large and slowly varying in the region \mbox{$|x-x_0| < \Delta
x$} and $S(x)$ goes through a few cycles there, then the time of
arrival distribution of the transmitted packet will be
approximately that of the incoming one, shifted by
$\frac{-L}{<v>}$. Note also that this is also true for a mixed
state which can be decomposed into various pure states with this
property.
\\Let us now find the explicit condition for $C(k)$ to be
approximately constant, for the case where \mbox{$L_j =
\frac{j}{n} L$}, $\alpha_j = \alpha$, and as before, $n < |\beta|
= \frac{m\alpha}{k}$. In this case, we have

\begin{equation} C(k) = \frac{\left( \frac{ik}{m\alpha}\right)^n}
{-2ik\left(2i \sin kL/n \right)^{n-1}}
\end{equation}
Using the fact that $\frac{x}{\sin x} = 1 + \frac{x^2}{6} +
O(x^4)$, we get:

\begin{equation} C(k) = -\frac{1}{m\alpha}\left(\frac{n-1}{2Lm\alpha}\right)^{n-1}
\left(1+ \frac{1}{6}(\frac{kL}{\sqrt{n-1}})^2 + O\left(\left(
\frac{kl}{n-1} \right)^3 \right) \right)
\end{equation}
Thus, $C(k)$ will be approximately constant if the spectrum of the
wave packet is limited to $k$ such that $|k| \approx
\frac{\sqrt{n-1}}{L}$. In other words, $\Delta k \approx
\frac{\sqrt{n-1}}{L}$, or $\Delta x \approx \frac{L}{\sqrt{n-1}}$.
This means that the length of the barrier can be arbitrarily
longer than the "length" of the tunneling packet as usually
defined (standard deviation of the $x$ coordinate), the penalty
paid being an exponential suppression of the amplitude. In passing
notice the function $F(k)$ in eqs. (1-2) displays a similar
behavior.

\subsection{Calculation of the dwell time}

It is interesting to compare the ``group delay'' (which is zero in
the low k limit) with the dwell time. For the sake of simplicity,
we shall deal with the case \mbox{$n = 2$}. A direct calculation
of the dwell time of the transmitted component can be made by
calculating the transmission coefficient after adding a potential
which is constant over the region between the delta spikes, and
vanishing outside it. We get:

\[ t \simeq \beta^{-1}\frac{e^{-ikL}}{-2i\sin k'L}, \]
where $k' = \sqrt{2m(E-V_0)/\hbar} $ and $V_0$ is the value of the
potential between the deltas. Clearly, $ \frac{\partial \arg
(t)}{\partial V_0} = 0 $, and the (conditional) dwell time is zero
as expected.

\section{Superluminality and its relation to interference in the tail of the wavefunction}
The calculation of the transmission coefficient, $t$,
 can also be done in a way more suggestive
of superoscillations. Let us explain this for the case of 2 delta
functions (the $n=2$ case in (\ref{potential}), ``Fabry Perot
interferometer'').

Suppose a quasimonochromatic wave packet with wave number $k$
arrives at the first delta spike. The transmitted component is the
same as the original wave, except for an attenuation and phase
which are independent of $k$. At the second delta spike, the wave
splits into a (approximately unattenuated) reflected wave and a
transmitted one which is apart from a $k$-independent
multiplicative constant the same as the impinging wave. The
reflected component is again reflected at the first delta, and
arrives at the second delta with an additional phase of $2kL$, but
with approximately the same amplitude as the original transmitted
wave. In a like manner, one gets additional transmitted waves with
additional phases of \mbox{$2nkL, n= 2,3,...$}, and amplitudes
which decrease very slowly. Thus we get the following formal sum
for the  resulting amplitude of the wave (up to a multiplicative
constant):

\begin{equation}
\sum_n e^{ikx}(e^{ik 2L})^{n} = \frac {e^{ikx}}{1-e^{2ikL}} =
e^{ikx} \frac{e^{-ikL}}{-2i\sin kL} \label{eq:superosc}
\end{equation}
which is in agreement with our previous calculation. This is an
example of superoscillations since a sum of positive wave vectors
rezults in a negative one (or, equivalently, a sum of positive
shifts which results in a negative shift).\footnote {The sum in
eq.(\ref{eq:superosc}) actually diverges, the physical reason
being that we have neglected the attenuation of the amplitude, in
order to maintain consistency with the low kinetic energy
approximation we have used so far. For the case $n=2$ it is easy
to evaluate eq.(\ref{tmatrix}) without resort to that
approximation, and the resulting transmission amplitude is:

\begin{equation} t(k) = \frac{\beta^{-2}}
{\left(1+\frac{2}{\beta}+\frac{1}{\beta^2}\right)-e^{2ikL}}
\end{equation}

Similarly, the sum on the left of eq.(\ref{eq:superosc}) should be
replaced by the exact one:
\begin{equation} (1+\beta)^{-2}\sum_{j=0}^\infty
\left[\left(\frac{\beta}{1+\beta}\right)^2 e^{2ikL}\right]^j =
\frac{\beta^{-2}}
{\left(1+\frac{2}{\beta}+\frac{1}{\beta^2}\right)-e^{2ikL}}
\end{equation}. }
This is true in the following sense: for $|k|<< 1/L$ the
denominator of the right hand side can be considered constant.
However, in such a small interval the function does not really
oscillate, so it really doesn't have a well defined frequency. To
really speak about superoscillations we need to have a large
number of delta-functions. The sum for the case $n>2$ factors into
$n-1$ sums of the above form, in the low kinetic energy limit,
since to leading order in $\beta^{-1}$, the only contributions are
from waves which tunnel through each delta only once, but may be
reflected any number of times between consecutive deltas. We then
reproduce the results of subsection 2.2, where we had the weaker
condition $k<\sqrt{n-1}/L$, allowing the function to complete many
oscillations in the region. Note the similarity to the situation
described by eqs. (1,2).

A different calculation of the (conditional) dwell time than that
of section 2.3 assumes a ``clock'' which is activated by the
presence of the particle in the region $[0,L]$, i.e.a degree of
freedom $\tau$ with an interaction Hamiltonian \mbox{$H_{\mathrm
{Int} } = \theta_{[0,L]}(x) p_{\tau}$}, where $p_\tau$ is the
canonical momentum conjugate to $\tau$. If the initial state of
the clock has small enough $\Delta p_\tau$, one gets for the final
phase of $\tau$ the same expression as eq. \ref{eq:superosc}, with
$x,k$ replaced by the clock's coordinates. This can be interpreted
as the sum appropriate for a weak measurement of
$\theta_{[0,L]}(x)$, as shown in the next section . Note that in
this case, not only do the dwell and phase times coincide, but
they are also described by the same mechanism.

The group delay in tunneling through a thick barrier follows from
the fact that under the barrier, no phase accumulates, and the
entire phase shift comes from the boundaries and is practically
independent of the thickness. For cases where interference with a
delayed wave, a few authors
\cite{CKS93,Stein94,Stein95c,Diener96} have suggested a
different explanation. In Chiao and Steinberg's words\cite{CS:rev}: "If
destructive interference is set up between part of the wave
travelling unimpeded and part which has suffered a delay $\Delta
t$ due to multiple reflections, one has
$\Psi_{out}(t)=\Psi_{in}(t)-\xi\Psi(t-\Delta t)\approx
(1-\xi)\Psi_{in}(t)+\xi\Delta td\Psi_{in}(t)/dt\approx
(1-\xi)\Psi_{in}(t+\xi\Delta t/(1-\xi))$, which is already a
linear extrapolation into the future. In cases where the
dispersion is sufficiently flat, as in a bandgap medium, the
extrapolation is in fact surprisingly better than this first-order
approximation. As was suggested by Steinberg \cite{Stein95c} 
and recently discussed more rigourously by Lee
and Lee \cite{2Lee} and Lee \cite{Lee96}, this implies that even a
simple Fabry-Perot interferometer exhibits superluminality when
excited off resonance" [presumably, $\xi << \frac{1}{\Delta t
\Psi'(t)}$ ]. We would like to explain this "better than
first-order" approximation. Let us instead look at the momentum
wave function. A spatial shift corresponds to a linear shift in
this function. A positive spatial delay would correspond to a
linear shift steeper than one, and the converse for a negative
delay. In the Taylor expansion of the transmission coefficient for
the momentum wave function, the zero term is insignificant, the
second corresponds, as just explained, to the spatial shift, and
the higher give the distortion. When many \emph{large} and evenly
distributed shifts waves interfere, their sum is for a wide range
of momenta, zero, and in particular momentum independent. In other
words, the momentum wave function is flat for a wide band of
frequencies. This corresponds to a much better than first order
approximation of the spatial wave function, as can be seem in the
special case of the system of section 3 of this paper.

\section{ The dwell time as a weak value}

We would like to calculate the expectation value of the time
measured by a "clock" consisting of an auxiliary system which
interacts weakly with our particle as long as it stays in a given
region. Furthermore, we would like to restrict the calculation
only to the subensemble of particles which ultimately end up on
the right of the barrier. The simplest interaction is perhaps the
one defined by the Hamiltonian:
\[ H_{\mbox{int}} = P_{\tau} X_{(0,L)} \]
Where $\tau$ is the degree of freedom of the clock and $P_\tau$ is
its conjugate momentum, and $X_{(0,L)}(x)=\left\{
\begin{array}{cc} 1 & \mbox{if $0<x<L$} \\
0 & \mbox{otherwise} \end{array} \right. $. This is the effective
form, for example of the potential , seen by a particle in an
$S_z$ eigenstate, in the Stern-Gerlach experiment ($\tau$ being
the $z$ coordinate, and $(0,L)$ the region of the magnetic field).
Assuming the clock to have at large negative times the expectation
value 0, a perturbation calculation shows that following the
interaction with the particle and the subsequent post-selection of
the particle state to be $|f\rangle$, the expectation value of
$\tau$ at large positive times is given by the formula:

\begin{equation} E(\tau, t \rightarrow \infty |i,f)=
\frac{\int_{-\infty}^{\infty}dt \int_{0}^{L} dx
\Psi_f^*(x,t)\Psi_i(x,t)} {\int_{\infty}^{\infty} dx
\Psi_f^*(x,0)\Psi_i(x,0)} \label{eq:condexp}
\end{equation}

Steinberg\cite{Steinberg1,Steinberg2} has arrived at this formula
under similar assumptions by a somewhat different line of
reasoning. He introduced the term conditional (quantum)
probability for the probability distribution of a system following
post-selection, and we use the notation of conditional expectation
in the formula above, in the same spirit. As noted by Steinberg,
the last equation 
is a special case of a weak value.

This formula is valid when $\tau$ and $p_\tau$ do not appear in
additional terms in the full Hamiltonian, but the generalization
is straightforward. To prove the formula, let us work in the
interaction picture. Denote the initial state of $\tau$ by
$|\phi_\tau(t)\rangle$ (and the corresponding wave function by
$\phi(\tau,t)$), and the initial (preselected) and final
(post-selected) states of the tunneling particle by
$|i(t)\rangle,|f(t)\rangle\ (\Psi_i(x,t),\Psi_f(x,t) )$,
respectively. Using the interaction picture and expanding to first order
in $P_\tau$\footnote{We can satisfy the condition of small perturbation by
choosing an initial clock wave function concentrated about $P_\tau$
= 0 with small enough uncertainty.}

\[
|\phi_\tau,t\rangle|i(t)\rangle= Te^{-{i\over\hbar}\int_{-\infty}^{t}
H_{I}(t')dt'} |\phi_\tau,t \rightarrow -\infty \rangle |i(t
\rightarrow -\infty)\rangle \simeq \]
\begin{equation}\left(1-{i\over\hbar}\int_{-\infty}^{t} H_{I}(t')dt'\right)
|\phi_\tau,t \rightarrow -\infty \rangle |i(t \rightarrow
-\infty)\rangle
\end{equation}where\[ H_I(t)=Te^{{i\over\hbar}\int_{-\infty}^{t} H_0(t')}
H_{\mbox{int}}e^{-{i\over\hbar}\int_{-\infty}^{t} H_0(t')}.\]After making the
post-selection of state $|f\rangle$ for the particle, the clock is
left in the state given by the above expression, multiplied on the
left by $ \frac{\langle f (t\rightarrow -\infty)|}{\langle
f,-\infty|i,-\infty \rangle}$. The expression we get after writing
out the explicit form of $H_{int}$ is:
\[ \phi(\tau,t\rightarrow +\infty) \simeq
e^{-i\langle X_{(0,L)} \rangle_W P_\tau/\hbar} \phi(\tau,t
\rightarrow -\infty) \] \begin{equation} =\phi(\tau + 
\langle X_{(0,L)}\rangle_W,t \rightarrow -\infty)
\end{equation} where
\begin{equation}
\langle X_{(0,L)} \rangle_W = \frac{\int_{-\infty}^{\infty}dt
\int_{0}^{L} dx \Psi_f^*(x,t)\Psi_i(x,t)} {\int_{\infty}^{\infty}
dx \Psi_f^*(x,0)\Psi_i(x,0)} \label{eq:condexp2}
\end{equation}
(the integral in the denominator is evaluated at $t=0$ merely for
convenience -- it is of course, time invariant). The wave
functions in the integral in the numerator can be taken in the
Schr\"{o}dinger representation. And the expectation for $\tau$ at
large positive times is shifted (from its initial value of 0) by
this value, as claimed. In contrast to the familiar eigenvalue
spectrum of a physical operator, its weak values can take any
complex values.\footnote{To see this, let us develop the initial
and final states of the particle in terms of the eigenfunctions of
the operator to be measured, $A$:
\begin{equation} |i> = \sum_k \alpha_{k} |a_{k}>, |f> = \sum_k \beta_{k} |a_{k}>,
(A|a_{k}\rangle = a_k|a_{k}\rangle)
\end{equation} Then, $A_W=\frac{\langle f|A|i\rangle}{\langle f|i\rangle}=\frac{\sum
\beta^*_k \alpha_k a_k }{\sum \beta^*_k \alpha_k} $ If $A$ is
nontrivial, it has more than one eigenvalue. Assume $k=1,2$
correspond to two of these, and take $\beta_1=\beta_2=
\frac{1}{\sqrt{2}}$ Then the two equations: $A_W= \frac{\alpha_1
a_1+\alpha_2 a_2}{\alpha_1+\alpha_2}= z$, and
$|\alpha_1|^2+|\alpha_2|^2=1$, are 3 real equations in 4 unknowns.
They can be solved for any value of $z$, as can be verified
easily. } 

\vspace {1cm }
The research was supported in part by grant 62/01-1 of 
the Israel Science Foundation, established by the
Israel Academy of Sciences and Humanities,
NSF grant PHY-9971005, and ONR  grant N00014-00-0383.



\eject


\begin{thebibliography}{99}


\bibitem{CS:rev} 
R.Y. Chiao and A.M. Steinberg, 1997, Progress in Optics vol.
XXXVII ,345.

\bibitem{Steinberg1} A.M. Steinberg, 1995,
Phys. Rev. A \textbf{52}, 32.

\bibitem{Steinberg2} A.M. Steinberg, 1995,
Phys. Rev. Lett. \textbf{74}, 2405.

\bibitem{YA:weak4}
Y. Aharonov, A. Casher, D. Albert, and
L. Vaidman. Phys. Lett. A124, 199 (1987).

\bibitem{aharonov}
Aharonov , D. Albert and L.
Vaidman, Phys. Rev. Lett.  60, 1351 (1988)

\bibitem{berry}
and Gaussian beams. M.V. Berry, 2000, J. Phys. A. \textbf{27}
L391, and {\em Faster than Fourier},  1994, Fundamental Problems
in Quantum Theory ed. J.A. Anandan and J. Safko.


\bibitem{ORS:2barr} V.S. Olkhovsky, E. Recami and G. Salesi, 2000,
arXiv:quant-ph/0002022 v3.
%


\bibitem{YA:weak1}
Y. Aharonov, P. G. Bergman and J.L. Lebowitz Phys. Rev. 134B, 1410-16
(1964)


\bibitem{YA:supero1}
Y. Aharonov, J. Anandan, S. Popescu and L.
Vaidman Phys. Rev. Lett. 64, 2965 (1990).

\bibitem{YA:weak6}
Y. Aharonov, C.K. Au and L. Vaidman, J. Phys. A: Math. Gen.  24, 2315
(1991)


\bibitem{YA:weak7}
B. Reznik and Y.
Aharonov. Phys. Rev A, 52, 2538 (1995)


\bibitem{CKS93} R.Y. Chiao, P.G. Kwiat and A.M. Steinberg,  1993,
Sci. Am. \textbf{269} 52.

\bibitem{Stein94} A.M. Steinberg, 1994, J. Phys. I (France)
\textbf{4}, 1813.

\bibitem{Stein95c} A.M. Steinberg, 1995, La Recherche
\textbf{281}, 46.

\bibitem{Diener96} G. Diener, 1996, Phys. Lett. A \textbf{223},
327.

\bibitem{2Lee} B. Lee, and W. Lee. 1995, Superlattices and
Microstructures \textbf{18}, 277.

\bibitem{Lee96} B. Lee, 1996, in: O.S.A. Annual Meeting
Abstracts, p. 185.



\end{thebibliography}
\end{document}